\documentclass[preprint,showpacs,preprintnumbers,floatfix]{revtex4}

\usepackage{epsfig} 
\usepackage{graphicx} 
\usepackage{dcolumn} 
\usepackage{bm} 

\newcommand{\beq}{\begin{eqnarray}}
\newcommand{\eeq}{\end{eqnarray}}
\newcommand{\lb}{\langle}
\newcommand{\rb}{\rangle}

\begin{document}
\title{Decoherence of a Josephson qubit due to coupling to two level systems}
\author{Li-Chung\ Ku and C.\ C.\ Yu}
\affiliation{Department of Physics, University of California,
    Irvine, California 92697, U.S.A.}
\date{\today}

\begin{abstract}
Noise and decoherence are major obstacles to the implementation of
Josephson junction qubits in quantum computing. Recent experiments suggest
that two level systems (TLS) in the oxide tunnel barrier are a source of
decoherence. We explore two decoherence mechanisms in which these two level
systems lead to the decay of Rabi oscillations that result when Josephson
junction qubits are subjected to strong microwave driving. (A) We consider
a Josephson qubit coupled resonantly to a two level system, i.e.,
the qubit and TLS have equal energy splittings. As a result of this
resonant interaction,
the occupation probability of the excited state of the qubit exhibits beating.
Decoherence of the qubit results when the two
level system decays from its excited state by emitting a phonon.
(B) Fluctuations of the two level systems in the oxide barrier
produce fluctuations and $1/f$ noise in the Josephson junction critical
current $I_o$. This in turn leads to fluctuations in the qubit energy
splitting that degrades the qubit coherence. We compare our results with
experiments on Josephson junction phase qubits.
\end{abstract}
\pacs{03.65.Yz, 03.67.Lx, 85.25.Cp}
\maketitle

\section{Introduction}

The Josephson junction qubit is a leading candidate as a basic component of
a quantum computer.  A significant advantage of this approach is scalability,
as these qubits may be readily fabricated in large numbers using
integrated-circuit technology.  Recent experiments on Josephson
qubits have successfully shown that they possess quantum coherent properties
\cite{Makhlin01,QB1,QB2,QB3,QB4,QB_flux,twoQB1,QB22}.
However, a major obstacle to the realization of
quantum computers with Josephson junction qubits is decoherence
and the dominant noise source has not yet been identified.
Because the measured decoherence times are substantially shorter than
what is needed for a quantum computer, there has been 
ongoing research to understand decoherence mechanisms
in Josephson qubits. Let us briefly review the theoretical work that
has been done.
Martinis et al.~\cite{QB3} investigated decoherence in 
Josephson phase qubits due to current noise primarily from external sources.
Paladino et al.~\cite{Paladino02} analyzed decoherence in 
charge qubits due to background charge fluctuations.
Van Harlingen et al.~\cite{Harlingen04,Harlingen04b} studied how low
frequency $1/f$ critical
current fluctuations lead to decoherence in various types 
of Josephson junctions. ($f$ denotes frequency.)
Smirnov \cite{Smirnov03} used perturbation theory to
study the decay of Rabi oscillations
due to thermal fluctuations of a heat bath weakly coupled to the qubit.
Even though fluctuating two level systems in oxide tunnel barriers 
have long been known to be a major intrinsic noise
source in Josephson junctions \cite{Wakai86,Wakai87}, their role in 
qubit decoherence has not been investigated theoretically. 
In this paper we will focus on the intrinsic microscopic mechanisms
whereby two level systems produce decoherence in a Josephson qubit. 

A recent experiment indicates that two level
systems can couple to a Josephson phase qubit \cite{Martinis04}.
In these experiments the lowest excitation frequency of a qubit is
measured as a function of the bias current that determines the
depth of the potential energy double well of the qubit. For most values
of the current bias, a single excitation frequency $\omega_{10}$ is
observed and it decreases with increasing bias current. Occasionally the
experiments find spurious resonances characterized by two closely
spaced excitation frequencies at a given bias current.
The size of the gap between the two excitation frequencies is on
the order of 25 MHz. Simmonds {\it et al.} \cite{Martinis04} have argued
that this splitting is evidence that the qubit is coupled to a
two level system (TLS) with an energy splitting very close or equal
to that of the qubit. We will refer to this as a resonant interaction.
The microscopic nature of these two level systems is unclear.
It may be due to the motion of oxygen atoms in the oxide tunnel
barrier of the Josephson junction \cite{Rogers85,Buhrman04}. An oxygen
atom (or vacancy) could sit in a double well potential and tunnel
between two positions. Or it may be due to a quasiparticle hopping
between two positions in the oxide barrier. Or an electron
trap in the oxide barrier could fluctuate between being occupied
and empty \cite{Rogers84,Rogers85,Wakai86,Wakai87}. Or a trapped
flux quantum could be tunneling back and forth between two positions
in the oxide barrier. Regardless of the microscopic nature
of the TLS, we can use the fact that it is a two level
system to understand how such a defect can
couple to the qubit. Note that the qubit energy splitting is
a function of the critical current $I_o$.
Fluctuations of the TLS lead to fluctuations
of the tunneling matrix element $T$ through the oxide barrier.
This in turn leads to fluctuations in
the critical current $I_o$ since $I_o\sim |T|^2$. If the two
states of the TLS correspond to two different values, $I_{o,1}$
and $I_{o,2}$, of the critical current, the coupling
between the qubit and the TLS is proportional to the
difference $|I_{o,1}-I_{o,2}|$.

Experiments often probe qubits using Rabi oscillations.
Let us take a moment to review Rabi oscillations \cite{Rabi}.
If the qubit is
initially in its ground state, resonant microwaves with
a frequency that matches the qubit energy splitting ($\omega_{10}$)
will initially increase
the probability amplitude of finding the qubit in its excited state
($|1\rangle$). However, as time goes on, at some point the qubit is
completely in its excited state, and the electromagnetic wave goes on
to de-excite the qubit through stimulated emission. Thus
the system will be coherently oscillating between the two energy eigenstates
with a Rabi frequency $f_R$.  The frequency $f_R$ of
the Rabi oscillations increases linearly with the amplitude of the
driving electric field.  Rabi oscillations have been seen
in the occupation probability $P_1$ of the excited state of
a Josephson qubit \cite{Rabi}. This demonstration of quantum
coherence is a
preliminary requirement for quantum computing but most of the reported Rabi
oscillations in Josephson qubits have a rather small amplitude (less
than 50\%) and a short coherence time (less than one microsecond).
Experiments have found that the presence of a resonant
interaction between the qubit and a two level system substantially
reduces or even eliminates Rabi oscillations \cite{Martinis04}.

In this paper we theoretically model a qubit coupled to a two level
system and study the effect of this coupling on Rabi oscillations.
We study the quantum dynamics of a Josephson qubit
by numerically integrating the time-dependent
Schr\"odinger equation. Our method is not limited to weak coupling to the noise
sources. While our analysis applies to any Josephson
qubit, for illustration we consider the Josephson phase qubit that was
studied in recent experiments \cite{Martinis04}.
We explore two decoherence mechanisms where two level systems
lead to the decay of Rabi oscillations that result when Josephson
junction qubits are subjected to strong microwave driving.
We first consider the resonant case.
In the resonant regime, the energy splitting of the two level
system and the qubit are matched. As a result
the occupation probability of the qubit's excited state exhibits beating.
This has been termed a qubit duet
\cite{twoQB1,QB22}. Our calculations show that decoherence
of the qubit results when the two level system decays from its excited
state by emitting a phonon. In section II, we numerically
calculate the qubit occupation probability as a
function of time and compare it with experiment.

The other case involves low frequency fluctuations of the qubit
energy splitting. Even though the
qubit energy splitting $\omega_{10}/2\pi$ is on the order 10 GHz,
low frequency fluctuations of the critical current lead to low
frequency fluctuations of the qubit energy splitting. We
hypothesize that this noise comes from slow fluctuations of
two level systems in the oxide tunnel barrier. If we assume
that there are a number of two level systems in the barrier,
and if these two level systems have a broad distribution
of decay rates, then they will produce $1/f$ noise
\cite{Dutta,Yu04} that leads to decoherence. 
In section III we show that these fluctuations
in the qubit energy splitting lead to decay of the Rabi oscillations.
We consider three cases. In the first case the qubit is coupled
to a single slowly fluctuating two level system. In the second
case the qubit is coupled resonantly to a two level system and,
at the same time, is subjected to slow fluctuations in the critical current
of a two level system. In the third case we consider a qubit with
energy splitting fluctuations that have a $1/f$ noise spectrum.

\section{Qubit-TLS Resonance}

In our model we assume that there are two level systems in
the oxide barrier of the Josephson junction.
The standard model of noninteracting two level
systems \cite{Hunklinger86,phillips} was introduced
by Anderson, Halperin, and Varma \cite{ahv}, and independently
by W. A. Phillips\cite{tls} in 1972.
The standard Hamiltonian for a two level system is
\begin{equation}
H=\frac{1}{2} \left( \begin{array}{cc}
\Delta & \Delta_o \\
-\Delta_o & -\Delta
\end{array} \right)\;.
\end{equation}
Here we are using the left well -- right well basis where $|L\rangle$
($|R\rangle$) is the left (right) well state.
$\Delta$ is the asymmetry energy, i.e., $\Delta$ is the
energy difference between the right well and the left well.
We can diagonalize the Hamiltonian to get the energy eigenvalues
that are given by $\pm \varepsilon_{TLS}/2$ where
\begin{equation}
\varepsilon_{TLS}=\sqrt{\Delta^2 + \Delta_o^2}\;.
\end{equation}

Simmonds {\it et al.} have found experimental evidence for resonant
interactions between a phase qubit and a two level
system \cite{Martinis04}.
This resonant interaction occurs when the energy splitting $\epsilon_{TLS}$
of the TLS matches that of the qubit, i.e.,
$\epsilon_{TLS} = \hbar \omega_{10}$. Simmonds {\it et al.} constructed
a phenomenological model to account for
their experimental findings \cite{Martinis04}.
In this section we investigate Rabi oscillations in the presence of a
qubit-TLS resonance. We incorporate decoherence into
the model of Simmonds {\it et al.} to show that Rabi oscillations
can decohere due to decay of the excited state of the two
level system via phonon emission.
We briefly describe the model of Simmonds {\it et al.} in the following.

The Hamiltonian of a Josephson phase qubit (which is essentially a
current-biased Josephson junction) is \cite{Leggett87,Martinis87,Clarke88}
\begin{eqnarray} H_{qb}  = \frac{{\hat Q^2 }}{{2C}} - \frac{{\Phi _0 I_o }}{{2\pi }}\cos
\hat \delta  - \frac{{\Phi _0 I_{bias} }}{{2\pi }}\hat \delta
\label{eq:QB_ham}
\end{eqnarray}
where $\Phi_0 = h/2e$ is a superconducting flux
quantum. $C$ is the capacitance of the junction, and $I_{bias}$ is
the bias current.
The operators $\hat Q$ and $\hat \delta$ correspond to the charge
and phase difference across the Josephson junction respectively.

Next we assume that there is a two level system in
the barrier of the Josephson junction. The two states of the TLS
correspond to two different values of 
the Josephson junction critical current $I_o$ which
is proportional to the square of the tunneling matrix element.
When the TLS is in state $|R \rangle$ (state $|L
\rangle$), the junction critical current is $I_{0R}$ ($I_{0L}$).
The qubit couples to the TLS because
the qubit's energy splitting $\hbar \omega_{10}$
is a function of $I_o$. The expression for $\omega_{10}$ can
be derived from the Hamiltonian in Eq.~(\ref{eq:QB_ham}) using
the expression for the resonant frequency of an $LC$ circuit
$\omega_{10} \approx 1/\sqrt{L_JC}$ where $L_J=\Phi_o/2\pi I_o\cos\delta$
is the Josephson inductance \cite{Martinis_acc}:
\begin{equation}
\omega _{10}  \approx \sqrt {\frac{{2\pi I_o }}{{\Phi _0 C}}} \left[
{2\left( {1 - \frac{{I_{bias} }}{{I_o }}} \right)} \right]^{1/4}.
\label{eq:w10}
\end{equation}
Here we used the fact that $I=I_o\sin\delta$ implies that
$\cos\delta=\sqrt{1-(I/I_o)^2}$. Typically $I_{bias}$ is slightly
less than $I_o$. The qubit couples to the TLS
because the Josephson junction critical current $I_o$ is
modified by the TLS. Therefore, the interaction Hamiltonian is
\cite{Martinis04}:
\begin{equation}
H_{qb - TLS}  =  - \frac{{\Phi _0 I_{0R} }}{{2\pi }}\cos \delta \otimes
\left| R \right\rangle \left\langle R \right| - \frac{{\Phi _0 I_{0L}
}}{{2\pi }}\cos \delta  \otimes \left| L \right\rangle \left\langle L
\right|.
\label{eq:H_int}
\end{equation}
We can transform to the eigenbasis of the TLS.
Provided that the TLS is symmetric, its
ground state is $|g \rangle = (|R \rangle + |L \rangle)/\sqrt{2}$
and its excited state is $|e \rangle = (|R \rangle - |L
\rangle)/\sqrt{2}$. The ground state of the qubit is $|0\rangle$
and the excited state of the qubit is $|1\rangle$. We can rewrite
the operator $\cos \hat \delta$ in terms of its matrix elements
in the qubit basis by noting that the phase qubit is typically biased close
to $\delta=\pi/2$ where the Josephson current is maximized.
So $\cos(\pi/2 - \delta^{\prime})=\sin(\delta^{\prime})\approx\delta^{\prime}$
where $\delta^{\prime}\ll 1$.
$\delta^{\prime}$ can be represented as a sum of a creation and an
annihilation operator in much the same way as the position coordinate
of a harmonic oscillator. Using these facts,
we find that Eq.~(\ref{eq:H_int}) becomes
\cite{Martinis04}
 \begin{eqnarray}
 H_{qb - TLS}  = \frac{{\delta I_o }}{2}\sqrt {\frac{\hbar }{{2\omega
_{10} C}}} \left( |0,g \rangle \langle 1,e |+ |1,e \rangle \langle 0,g |+
|1,g \rangle \langle 0,e | + |0,e \rangle \langle 1,g | \right),
 \end{eqnarray}
where $\omega_{10}$ is the energy difference in the qubit levels,
and $\delta I_o \equiv I_{0R}-I_{0L}$ is the
fluctuation amplitude in $I_o$ produced by the TLS.
Since the values of $\langle H_{qb-TLS} \rangle$, $C$ and $\omega_{10}$
can be determined experimentally,
one can estimate $\delta I_o/I_o$ to be approximately $6\times 10^{-5}$
\cite{Martinis04} when the Josephson junction is in the
zero-voltage state. However this value of $\delta I_o/I_o$
differs from previous measurements made on Josephson junctions
in the finite voltage state that found low frequency (1 kHz
or less) fluctuations with $\delta I_o/I_o\sim 2\times 10^{-6}$
\cite{Harlingen04,Wakai86}. 

Next we study the quantum dynamics of the coupled qubit-TLS system when it
is subjected to microwave driving at a frequency $\omega_{10}$ equal
to the qubit energy splitting.
We will assume that both the qubit and the TLS can couple to
microwaves. The resulting Hamiltonian matrix of the qubit-TLS model is,
\begin{eqnarray}
H_{qb-TLS} = \left( {\begin{array}{*{20}c}
{\mbox{\ \ }0 \mbox{\ \ }} & {\mbox{\ \ \ }{g_{TLS} \sin(\omega_{10}t)
}\mbox{\ \ \ }} &
{\mbox{\ \ \ }{g_{qb}\sin(\omega_{10}t) }\mbox{\ \ \ }} & \eta  \\
{g_{TLS}\sin(\omega_{10}t) } & {\varepsilon _{TLS} } & \eta &
{g_{qb}\sin(\omega_{10}t) }  \\
{g_{qb}\sin(\omega_{10}t) } & \eta & \hbar {\omega _{10} } & {g_{TLS}
\sin(\omega_{10}t)}  \\
\eta & {g_{qb}\sin(\omega_{10}t) } & {g_{TLS} \sin(\omega_{10}t)} & {\hbar
\omega _{10}
   + \varepsilon _{TLS} }  \\
\end{array}} \right) \label{eq:ham}
 \end{eqnarray}
where the basis states are $|0, g\rangle$, $|0, e\rangle$, $|1, g\rangle$,
and $|1, e\rangle$ respectively. $\eta=(\delta I_o /2)\sqrt
{\hbar/(2\omega _{10} C)}$ is the coupling between the qubit and
the TLS; $g_{qb}$ is the coupling between the qubit and the microwaves;
and $g_{TLS}$ is
the coupling between the TLS and the microwaves. In the strong driving
regime, the qubit-microwave coupling is larger than the qubit-TLS
coupling. We consider the case of strong driving throughout the paper
because it is the experimental condition in Ref.~\cite{Martinis04}.
Without microwave driving, the Hamiltonian matrix can be
decoupled into two $2 \times 2$ matrices. We calculate the time evolution of
the qubit-TLS wave function by integrating the time-dependent Schr\"odinger
equation using the Runga-Kutta method\cite{Kutta}.
The initial wave function $|\psi\rangle$ is the
ground state of Eq.~(\ref{eq:ham}). When $\hbar \omega_{10}$ and
$\varepsilon_{TLS}$ are much greater than $\eta$, the ground state wave
function is approximately given by $|0, g \rangle$.

\begin{figure}
\centerline{\psfig{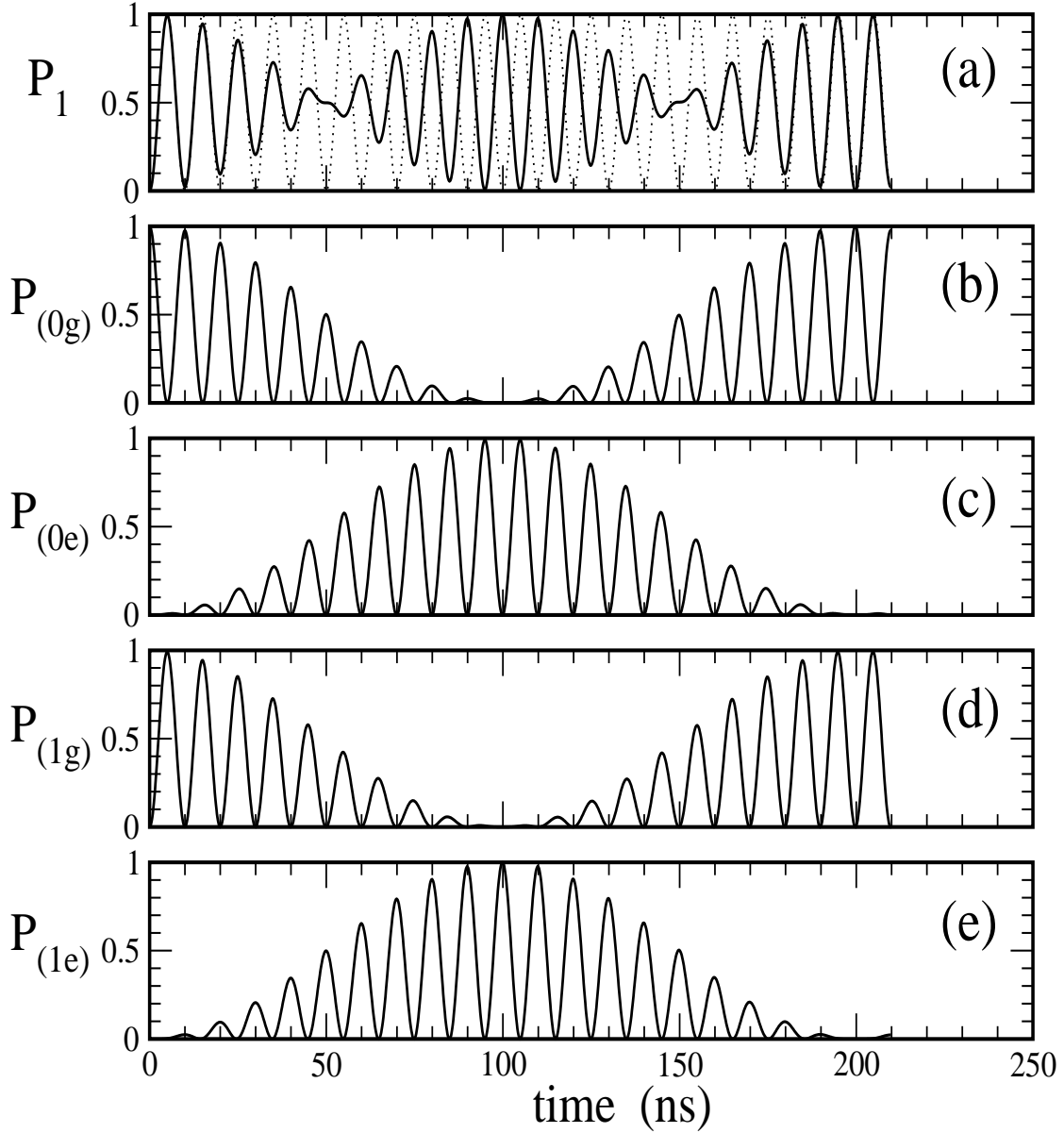}}
\caption{Rabi oscillations of a resonantly coupled qubit-TLS system with
$\varepsilon_{TLS} = \hbar \omega_{10}$.
There is no mechanism for energy decay.
Occupation probabilities of various states
are plotted as functions of time. (a) $P_1$ is the
occupation probability in the qubit state $|1\rb$;
(b) $P_{(0,g)}$ is the occupation
probability in the state $|0,g\rb$; (c) $P_{(0,e)}$
is the occupation probability of the state $|0,e\rangle$;
(d) $P_{(1g)}$ is the occupation probability of the state $|1,g\rangle$;
and (e) $P_{(1e)}$ is the occupation probability of the
state $|1,e\rangle$. Notice the beating with frequency
$2\eta$. Throughout the paper,
$\omega_{10}/2\pi = 10$ GHz. Parameters are chosen mainly according to the
experiment in Ref.~\cite{Martinis04}: $\eta/\hbar \omega_{10} = 0.0005 $,
$g_{qb}/\hbar \omega_{10} = 0.01 $, and $g_{TLS} = 0$. The dotted line in
panel (a) shows the usual Rabi oscillations without resonant interaction,
i.e. $\eta= 0$.  }
\label{fig:QB4L}
\end{figure}

We first consider the case of strong driving with $g_{TLS}=0$ and with
the TLS in resonance with the qubit, i.e.
$\varepsilon_{TLS} = \hbar \omega_{10}$.
If there is no coupling between the qubit and the TLS, then the four states
of the system are the ground state $|0,g\rangle$, the highest energy state
$|1,e\rangle$, and two degenerate states in the middle $|1,g\rangle$ and
$|0,e\rangle$. If the qubit and the TLS are coupled with
coupling strength $\eta$, the degeneracy is split by an energy $2\eta$.
Figure \ref{fig:QB4L} shows the coherent oscillations of the
resonant qubit-TLS system. We define a projection operator
$\hat P_1 \equiv |1, g\rangle \langle 1, g | + |1, e\rangle \langle 1, e | $
so that $\langle \hat P_1 \rangle$ corresponds to the
occupation probability of the qubit to be
in state $|1 \rangle$ as in the phase-qubit experiment. Instead of being
sinusoidal like typical Rabi oscillations (the dotted curve), the occupation
probability $P_1$ exhibits beating (Fig.~1a) because the two entangled
states that are linear combinations of $|1,g\rangle$ and
$|0,e\rangle$ have a small energy splitting
$2\eta$, and this small splitting is the beat frequency. Without
any source of decoherence,
the resonant beating will not decay. Thus far the
beating phenomenon has not yet been experimentally verified. The
lack of experimental observation of beating implies that the TLS or
qubit decoheres in less time than the period $\approx 1/\eta$. Note in
Fig.~\ref{fig:QB4L}a that the second beat is out of phase when compared with
the usual Rabi oscillations. The occupation probabilities in the individual
states are plotted in Fig.~1b-1e. (To the best of
our knowledge, these quantities are
not measurable.) We find that there is a very low occupation probability in
the states $|0, e\rangle$ and $|1, e\rangle$
during the first Rabi cycle because the
TLS is not directly coupled to the microwaves, and thus the TLS tends to
be in its ground state $|g\rangle$. In the limit $\eta \to 0$, the system
oscillates coherently between $|0,g\rangle$ and $|1,g \rangle$.
Occupying the two states $|0, e\rangle$ and $|1, e\rangle$ occurs only
via the qubit-TLS resonance coupling $\eta$. From Fermi's golden rule,
the average transition rate to $|0,e\rangle$ from $|1,g \rangle$ is
$2\pi\eta^{2}/\hbar$ which is much slower than the initial transition
rate from $|0,g\rangle$ to $|1,g\rangle$.

Now we consider the decay of excited TLS via phonons as a
source of decoherence for the qubit-TLS system. Two level systems
couple to the strain field and are able to decay from their excited
state by emitting a phonon with an energy equal to the TLS
energy splitting. The rate for an
excited TLS to emit a phonon and return to its ground state is given by
\cite{Hunklinger86}
\begin{equation}
\tau _{ph}^{ - 1}  = \frac{{\tilde \gamma ^2 }}{\rho
}\left( {\frac{1}{{c_l^5 }} + \frac{2}{{c_t^5 }}} \right)\frac{{
\varepsilon_{TLS} ^3 }}{{2\pi \hbar ^4 }}\left( {\frac{{\Delta _0 }}{{
\varepsilon_{TLS} }}} \right)^2 \coth \left( {\frac{{\beta \varepsilon_{TLS}
}}{2}} \right)
\label{eq:tau}
\end{equation}
where $\tilde \gamma$ is the deformation potential, $\rho$ is the mass
density, $c_l$~($c_t$) is the longitudinal (transverse) speed of sound,
and $\beta$ is the inverse temperature.
From Eq.~(\ref{eq:tau}), the relaxation time $\tau_{ph}$ is estimated to
be in the range from 10 to 100 ns. (We find $\tau_{ph} \sim 80$ ns when using
the following values that are appropriate for a symmetric
TLS in SiO$_2$ with a tunnel splitting that matches the energy splitting of
a Josephson junction qubit: $\tilde \gamma = 1.0$ eV,
$\rho = 2.2 $ g/cm$^3$, $c_l= 5.8 \times 10^{5}$ cm/s,
$c_t=3.8\times 10^{5}$ cm/s, $\Delta=0$, $\Delta_o=0.5 K$,
and $T =25$ mK.)

The decay of the excited two level system leads to decoherence of
the Rabi oscillations.
We can incorporate this relaxation rate of the excited TLS
into our calculations
of the Rabi oscillations of the qubit-TLS system by using the Monte
Carlo wave function method \cite{Monte}. In the original
application \cite{Monte} a two-level atom driven by a laser
field can decay by emitting a photon. We can easily
generalize the algorithm for our somewhat more complicated case
in which the qubit-TLS system decays
either from $|0, e \rangle$ to $|0, g \rangle$ or
from $|1, e \rangle$ to $|1, g \rangle$.
The algorithm goes
as follows. (a) Numerically propagate the wave function from time $t_i$ to
$t_i + \Delta t$ as if there is no energy decay in the TLS. Note that the
probability for decaying from $| 0,e \rangle$ to $| 0,g \rangle$ during the
time interval $\Delta t$ is $P_{(0,e)}\times \left[1- {\exp \left( {\frac{{
- \Delta t}}{{\tau _{ph} }}} \right)} \right]$, where $P_{(0,e)}$ is
the probability that state $|0,e\rangle$ is occupied. (b) Generate a uniformly
distributed random number $r_i \in [0,1]$.
If $ r_i < P_{(0,e)}\times  \left[1- {\exp \left( {\frac{{ - \Delta
t}}{{\tau _{ph} }}} \right)} \right]$, then the system
decays from $|0,e \rb $ to
$|0,g \rb $, and we represent this by
resetting the wave function $\psi(t_i + \Delta t)$ to $ |
0, g \rangle$. Then we repeat steps (a) and (b). Similarly, if $ 1- r_i <
P_{(1,e)}\times \left[1- {\exp \left( {\frac{{ - \Delta t}}{{\tau _{ph} }}}
\right)} \right]$, then the system decays from $|1,e\rangle$ to
$|1,g\rangle$, and we represent this by resetting the wave function
$\psi(t_i + \Delta t)$ to $ |1, g\rangle$. Then we repeat steps (a)-(b).
Notice that $\left[1- {\exp \left( {\frac{{ - \Delta t}}{{\tau _{ph} }}}
\right)} \right]$ is very small since the time step $\Delta t$ is small.
So there is no chance that both decays could happen in the same time step.
If neither of the above criteria are satisfied, then the qubit does not decay.
Then we repeat steps (a)-(b)
and keep propagating the wave function until the desired finishing time.

\begin{figure}
\centerline{\psfig{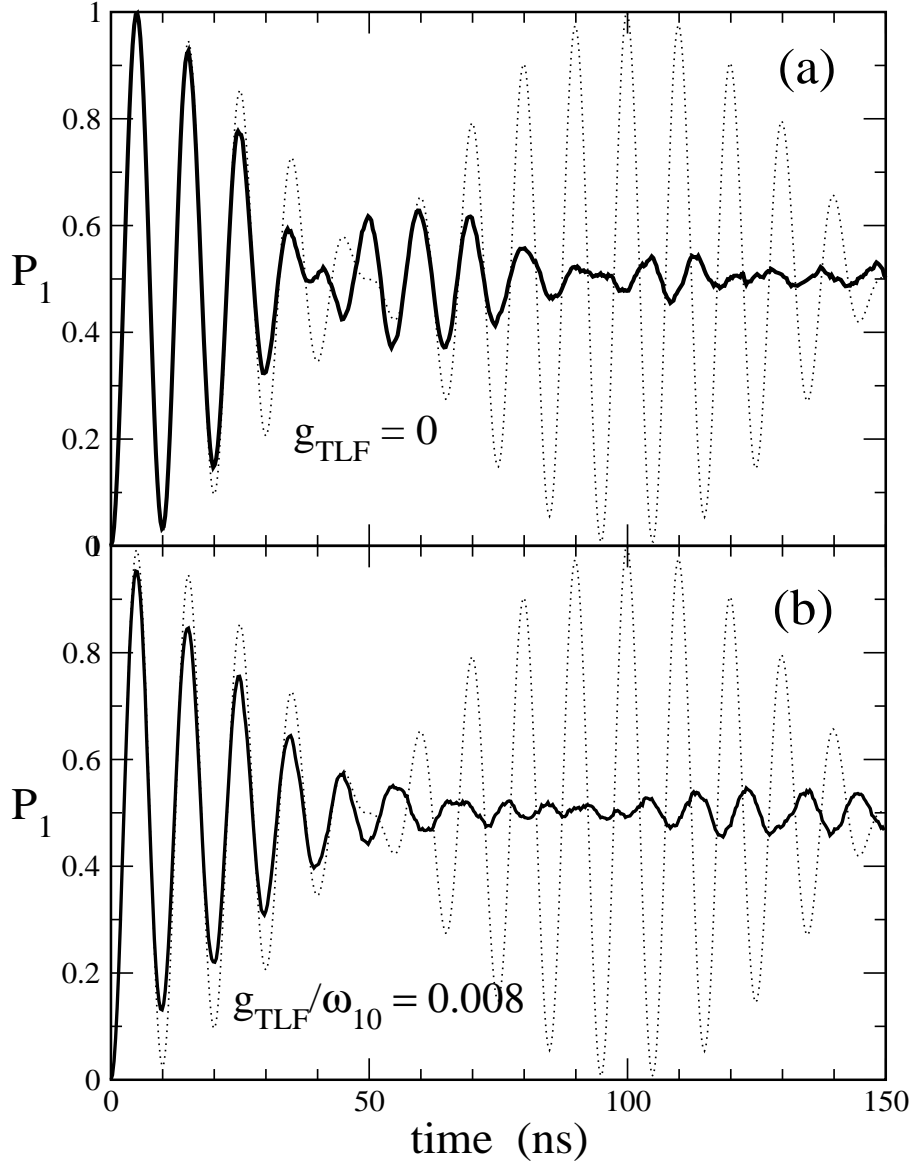}}
\caption{Relaxation of Rabi oscillations due to energy decay of the TLS
via phonon emission.
Solid lines include the effect of the energy decay of 
the TLS with $\tau_{ph} = 10$
ns and the dotted lines have $\tau_{ph} =\infty$.
These results are averaged over 100 runs in this figure and
in all of the following figures. 
(a) No coupling between the TLS and the microwaves.
(b) Direct coupling between the TLS and the microwaves with
$g_{TLF}/\omega_{10}=0.008$.
The rest of the parameters are the same as in Fig.1. We note that the
coupling between the TLS and microwaves degrades the qubit coherence.
}
\label{fig:10ns} 
\end{figure}

We have used our algorithm to study the effect of TLS energy decay
via phonon emission on Rabi oscillations.
The results are shown in Fig.~\ref{fig:10ns}.
The dotted lines show the beating that occurs when the qubit and TLS
are in resonance with no TLS decay. The solid lines shows the rapid damping
of the Rabi oscillations and the dephasing that occurs
when the TLS can decay via phonon emission. In Fig.~\ref{fig:10ns}a
there is no direct coupling between the microwaves and the TLS,
whereas in Fig.~\ref{fig:10ns}b, the microwaves are directly coupled
both to the qubit and to the TLS. We see that in the latter case beating
is damped out more quickly.
The Rabi decay time $\tau_{Rabi}$ can be defined as the time for
the envelope of the Rabi oscillations to decay by 1/e of their
original amplitude. Fig.~\ref{fig:10ns} shows that $\tau_{Rabi}$
is longer than
$\tau_{ph}$ simply because the excited state of the TLS is not always
occupied and available for decay.
Fig.~\ref{fig:QB4L} shows that states $|0,e
\rangle$ and $|1, e \rangle$ are essentially unoccupied at early times
and that the TLS energy decay can take place only when the
excited state of the TLS is sufficiently populated. Therefore in
Fig.~\ref{fig:10ns}a the solid and dotted lines are in phase
for the first few Rabi cycles. 

Figure \ref{fig:10ns}b shows the effect of coupling between the
microwaves and the TLS.  It has been shown experimentally
that a TLS will couple to microwaves
if it has an intrinsic or induced dipole moment \cite{Golding79,Golding80}.
We can estimate the value of the matrix element $g_{TLS}$ to be approximately
$pE$ where $p$ is the magnitude of the dipole moment and $E$ is
the magnitude of the electric field produced by the incident microwaves.
We can estimate $p$ by assuming that
the TLS is a charged particle (with charge $|q|=e$)
hopping between two sites separated
by a couple of angstroms. We can estimate $E \sim V/d$ where $d\sim 20 \AA$ is
the thickness of the oxide barrier and $V$ is the voltage produced across
the junction by microwaves. We can calculate $V$ using the Josephson equations
with the current being the microwave current $I_{\mu w}$
across the junction. Using the experimental values from 
Ref.~\cite{Martinis04}, we can estimate $I_{\mu w} \approx 1$ pA.
Therefore, we have $g_{TLS} \approx 0.01\hbar \omega_{10}$, which is
comparable to $g_{qb}$.
Comparing the two panels, we find that the TLS-microwave coupling
greatly enhances the resonant decoherence mechanism because
the transition rates from the ground state to the
states $|0,e\rangle$ and
$|1,e\rangle$ increase as $g_{TLS}$ increases. We conclude that the energy
decay of the TLS mainly causes the Rabi oscillations to decay,
and adding the TLS-microwave coupling further degrades the qubit
coherence. However, the phase qubit experiment indicates that the resonant
interaction affects both the Rabi amplitude and decay time
\cite{Martinis_pr}. The large amplitude of the Rabi oscillations that
we see at short times is not seen experimentally. Experimentally,
a qubit in resonance with a TLS has very small amplitude Rabi
oscillations at all times. This implies that our calculations do
not include all
the sources of decoherence responsible for the
experimental observations, such as resonance with
multiple fluctuators and interactions between the fluctuators.

\begin{figure}
\centerline{\psfig{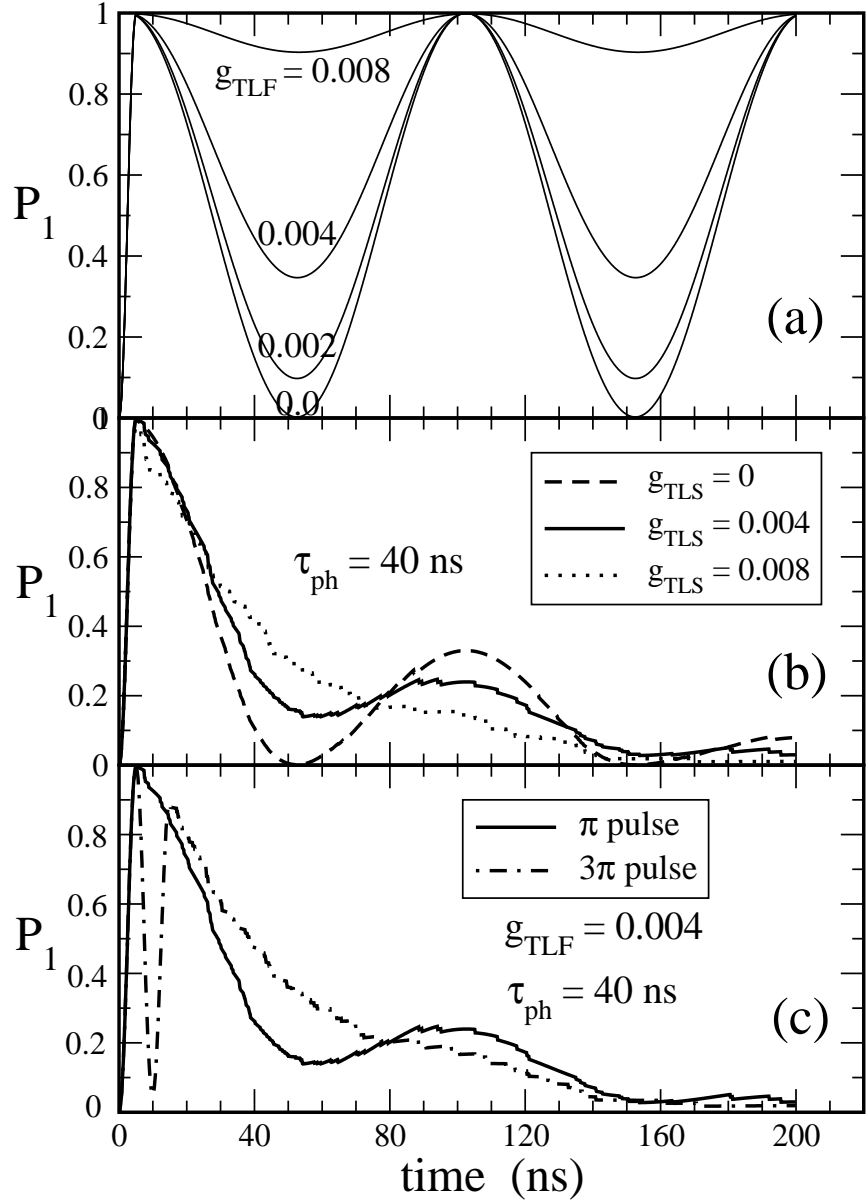}}
\caption{The qubit-TLS system starts in its ground state at $t=0$.
A microwave $\pi$ or $3\pi$ pulse (from 0 to 5 ns) puts
the qubit-TLS system in the qubit excited state $|1\rangle$ that
is a superposition of the two entangled states $|\psi'_1 \rangle \equiv (|0,
e\rangle + |1, g\rangle)/\sqrt{2}$ and
$|\psi'_2\rangle \equiv (|0, e\rangle - |1, g\rangle)/\sqrt{2}$.
After the microwaves are turned off,
the occupation probability starts oscillating coherently. Values
of $g_{TLS}$ indicated in the figure are normalized by $\hbar \omega_{10}$.
The rest of the parameters are the same as in Fig. 1. (a)
No energy decay of the excited TLS, i.e., $ \tau_{ph} = \infty$.
Coherent oscillations with various values of $g_{TLS}$. (b)
Oscillations following a $\pi$-pulse with $ \tau_{ph} = 40$ ns
and various values of $g_{TLS}$. (c) Oscillations following a $\pi$-pulse
and a $3\pi$-pulse with $\tau_{ph} = 40$ ns and $g_{TLS}=0.004$.
The dip in the dot-dash line is one and a half Rabi cycles.
}\label{fig:Rabi_pi} 
\end{figure}

Next we propose an experiment to test if the TLS in the Josephson
qubit couples to microwaves or not.
Start in the ground state, then send in a resonant
microwave $\pi$-pulse and stop
pumping. (A $\pi$-pulse lasts for half of a Rabi cycle.) When $g_{TLS}
\ll g_{qb}$, the qubit-TLS system mainly occupies the state $|1, g \rangle $
right after the $\pi$ pulse. State $|1, g \rangle $ is a superposition of
two eigenstates ($|\psi'_1 \rangle = (|0, e\rangle + |1, g\rangle)/\sqrt{2}$
and $|\psi'_2 \rangle = (|0, e\rangle - |1, g\rangle)/\sqrt{2}$)
of the qubit-TLS system.  Thus the wave function oscillates between
the states $|0, e \rangle$ and $|1, g \rangle$ after the $\pi$-pulse.
Provided that the relaxation time of the TLS
is longer than $1/\eta$, coherent oscillations in the
occupation probability $P_1$ can be observed with a
period of $2\eta$ \cite{KenCooper}. If there is no
energy decay of the excited TLS, there is no mechanism for decoherence and
the oscillation amplitude is
one. When $g_{TLS}$ is comparable to $g_{qb}$, both states $|1, g\rangle $
and $|1, e\rangle $ are partially occupied right after the $\pi$-pulse.
State $|1, e\rangle $ is essentially an stationary state. Thus partially
occupying $|1, e\rangle $ reduces the Rabi oscillation
amplitude of $P_1$ and merely gives a constant contribution to $P_1$. Figure
\ref{fig:Rabi_pi} shows the dynamics of the qubit-TLS system during and
after a microwave $\pi$-pulse. As we expect,
Fig.~\ref{fig:Rabi_pi}a shows that the oscillation amplitude decreases as
$g_{TLS}$ increases. In Fig.~\ref{fig:Rabi_pi}a the
excited TLS has no means of decay ($\tau_{ph}=\infty$), so
the coherent oscillations do not
decay. In Fig.~\ref{fig:Rabi_pi}b and Fig.~\ref{fig:Rabi_pi}c,
we set $\tau_{ph}=$ 40 ns. As a result, the
oscillatory $P_1$ is attenuated. We include the effect of the
energy decay of the TLS by using the
Monte Carlo wave function method described previously.
These model calculations suggest that one can estimate $g_{TLS}$ by
measuring the oscillation amplitude. An alternative way to measure
$g_{TLS}$ is to compare the Rabi oscillations after
$\pi$ and $3\pi$ pulses. If $g_{TLS}=0$, right after a $\pi$ or $3\pi$ pulse,
the system primarily occupies the state $|1,g\rangle$ with $P_{(1,g)}\sim 1$.
Therefore the coherent oscillations
after the $\pi$ and 3$\pi$ pulses ought to be similar. However, if $g_{TLS}$
is nonzero, the longer pulse will pump more weight into state $|0,e \rb$.
So the amplitude of the oscillations of $P_1$ right after 
a $\pi$-pulse will be greater than immediately after a $3\pi$-pulse. With
a 3$\pi$-pulse, the oscillations in $P_1$ are more damped, as shown
in Fig.~\ref{fig:Rabi_pi}c.

\section{The low-frequency two-level fluctuator}

As described in Sec.~I, a two level system trapped inside a Josephson junction
barrier can produce noise in the critical current $I_o$ by varying the height
of the tunneling potential barrier. As a result, the qubit energy levels
fluctuate and this leads to decoherence. We now study the
decoherence produced by a low-frequency TLS. As we mentioned earlier,
the dependence of $\omega_{10}$ on $I_o$ is given by
Eq.~\ref{eq:w10}. We note that $\omega
_{10}$ is modulated as $I_o$ varies. As the TLS modulates $I_o$, the phase
of the qubit is randomized and coherent temporal oscillations are destroyed.
Using the fact that $I_{bias}$ is slightly smaller than $I_o$
in Eq.~(\ref{eq:w10}), one finds
\begin{eqnarray} \frac{{\delta \omega _{10} }}{{\left\langle {\omega _{10} }
\right\rangle }} \approx \frac{{\delta I_o }}{{4\left( {\left\langle {I_o }
\right\rangle  - I_{bias} } \right)  }}~. 
\label{eq:w10_I0} \end{eqnarray}
In Ref.~\cite{Martinis04}, it was found that 
$\delta I/\langle I_o \rangle \approx 6 \times 10^{-5}$.
In addition, the phase qubit is typically operated at a bias current such
that $(\langle I_o \rangle - I_{bias})/ \langle I_o \rangle= 0.0025$.
Substituting these numbers into Eq.~(\ref{eq:w10_I0}), one can estimate that
the amplitude of the level fluctuations $\delta \omega_{10} / \langle
\omega_{10} \rangle $ is approximately 0.006.

\begin{figure}
\centerline{\psfig{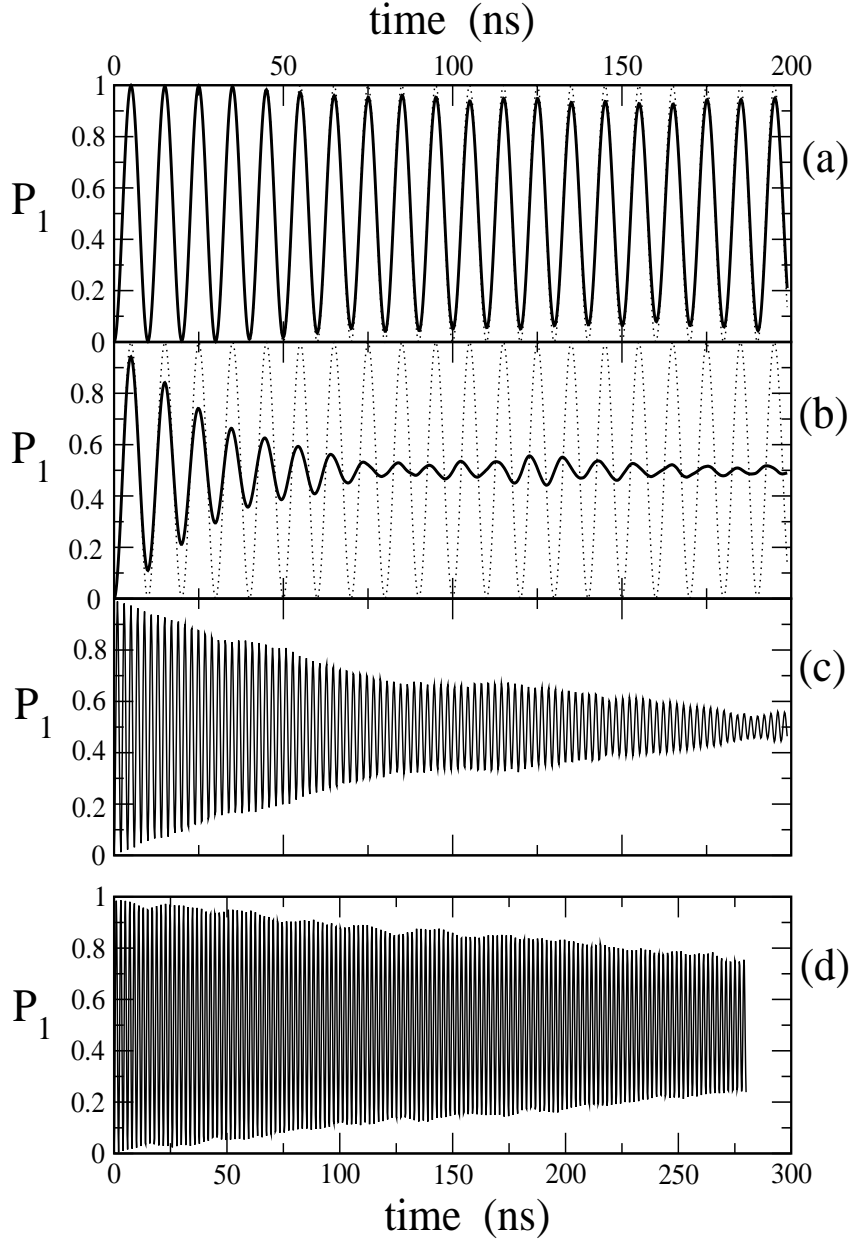}}
\caption{Solid lines show the Rabi oscillation decay due to qubit level
fluctuations caused by a single fluctuating two level system trapped
inside the insulating tunnel barrier.
The TLS produces random telegraph noise in $I_o$ that modulates the qubit energy
level splitting $\omega_{10}$. (a) The level fluctuation
$\delta \omega_{10}/ \langle \omega_{10} \rangle = 0.001$.
The characteristic fluctuation rate $t_{TLS}^{-1} = 0.6$ GHz.
The Rabi frequency $f_{R} = 0.1$ GHz.
The dotted lines show the usual Rabi oscillations
without any noise source. (b) $\delta \omega_{10}/ \langle
\omega_{10} \rangle = 0.006$, $t_{TLS}^{-1} = 0.6$ GHz, and
$f_{R} = 0.1$ GHz.
The dotted lines show the usual Rabi oscillations
without any noise source. (c) $\delta \omega_{10}/ \langle
\omega_{10} \rangle = 0.006$, $t_{TLS}^{-1} = 0.6$ GHz, and
$f_{R} =0.5$ GHz. (d) $\delta \omega_{10}/ \langle
\omega_{10} \rangle = 0.006$, $t_{TLS}^{-1} = 0.06$ GHz, and
$f_{R} =0.5$ GHz. Note that the scales of the horizontal axes
in (a)-(c) are the same. They are different from that in (d).
}\label{fig:RTN} 
\end{figure}

The Rabi oscillations are calculated by
integrating Schr\"odinger's equation in the presence of noise from a single
low-frequency fluctuator, i.e., $\omega_{10}t_{TLS} \ll 1$. $t_{TLS}$ is
the characteristic time of the random switching of the TLS.
We simulate the noise using the Monte Carlo method. At each time
$t_i$, a random number $r_i \in [0,1] $ is generated. If
$r_i < ((\mbox{time step})/t_{TLS}) < 1$, then the two level system
switches wells. We can imagine that these are thermally activated
transitions between the two wells of a symmetric double well potential.
We assume here that the dwell times $t_{TLS}$ are the same in the two
wells. Each time the TLS switches wells, the critical current, and hence
$\omega_{10}$ switch between two values.
The Rabi oscillations of the qubit are calculated using
the Hamiltonian
\begin{eqnarray} H(t) = \left( {\begin{array}{*{20}c}
   0 & {g_{qb} \sin (\omega _{10} t)}  \\
   {g_{qb} \sin (\omega _{10} t)} &
   \hbar [{\omega _{10}  + \delta \omega _{10}(t) }]  \\
\end{array}} \right) \label{eq:RTN} \end{eqnarray}
where the qubit energy levels ($|0\rb$ and $|1 \rb$) are the basis states
and the noise is produced by a single TLS.
Our calculations are oriented to the experimental conditions and the results
are shown in Fig.~\ref{fig:RTN}. In Fig.~\ref{fig:RTN}a-c
the characteristic fluctuation rate $t_{TLS}^{-1} = 0.6$ GHz. 
Panel \ref{fig:RTN}a shows that the qubit
essentially stays coherent when the level fluctuations are small ($\delta
\omega_{10}/\omega_{10} =0.001$). Panel \ref{fig:RTN}a shows that
when the level fluctuations increase to
0.006, the Rabi oscillations decay within 100 ns. The Rabi relaxation
time also depends on the Rabi frequency as panel \ref{fig:RTN}c shows.
The faster the Rabi
oscillations, the longer they last. This is because the low-frequency noise 
is essentially constant over several
rapid Rabi oscillations \cite{Smirnov03}. Alternatively, one
can explain it by the noise power spectrum $S_I(f)$. Since the noise
from a single TLS is a random process characterized by a single
characteristic time scale $t_{TLS}$, it has a Lorentzian power spectrum
\cite{S1_1,S1_2,S1_3,S1_4}
 \begin{eqnarray}
S_I (f) \sim \frac{{t_{TLS} }}{{(2\pi ft_{TLS})^2 + 1}}
\label{eq:Lorentzian}
\end{eqnarray}
We do not
expect Rabi oscillations to be sensitive to noise at frequencies
much greater than the frequency of the Rabi oscillations
because the higher the frequency $f$, the smaller the noise power and
because the Rabi oscillations will tend to average over the noise.
Rabi dynamics are sensitive to the noise at frequencies
comparable to the Rabi frequency.
In addition, the
characteristic fluctuation rate plays an important role in the rate of
relaxation of the Rabi oscillations. It has been
shown that $t_{TLS}^{-1}$ can be thermally activated \cite{noise_T} for 
TLS in a metal-insulator-metal tunnel junction. If the thermally activated
behavior applies here, the decoherence time $\tau_{Rabi}$ should decrease
as temperature increases.
In Fig.~\ref{fig:RTN}d, the characteristic fluctuation rate has
been lowered to 0.06 GHz
(which is much lower than $\omega_{10}/2\pi \approx 10 $ GHz).
The noise still causes qubit decoherence but affects the qubit
less than in Fig.~\ref{fig:RTN}c. Fig.~\ref{fig:RTN} shows
that the noise primarily affects the Rabi amplitude rather than the phase.

\begin{figure}
\centerline{\psfig{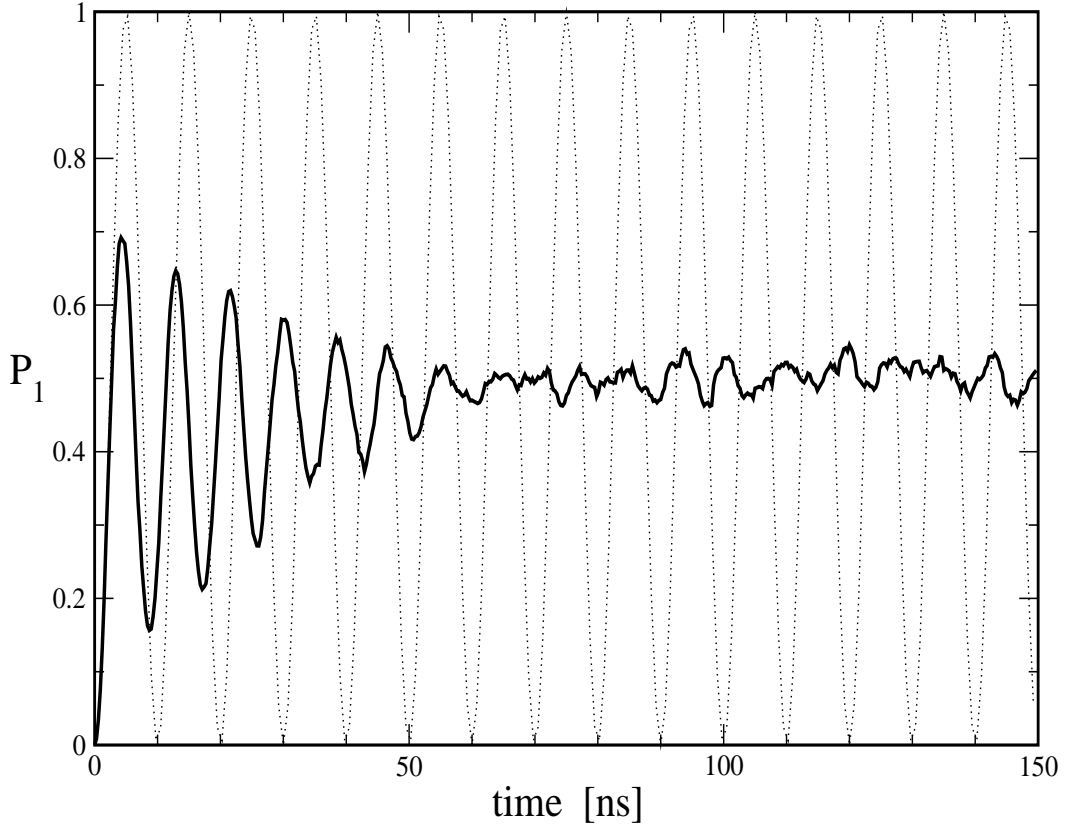}}
\caption{Solid line represents Rabi oscillations in the presence of both TLS
decoherence mechanisms: resonant interaction between the TLS and the qubit,
and low frequency qubit energy level fluctuations caused by a single
fluctuating TLS. The TLS couples to microwaves ($g_{TLS}/(\hbar
\omega_{10}) = 0.008$) and the energy decay time for the TLS is
$\tau_{ph}=10$ ns, the same as in Fig.~2b. The size of the qubit level
fluctuations is the same as in Fig.~\ref{fig:RTN}b. The dotted line shows the
unperturbed Rabi oscillations.}\label{fig:RTN_decay} 
\end{figure}

Experimentally, the two TLS decoherence mechanisms (resonant interaction and
low-frequency level fluctuations) can both be active at the same time.
We have calculated the Rabi oscillations in the presence of both of these
decoherence sources by using the qubit-TLS Hamiltonian in
eq.~(\ref{eq:ham}) with a fluctuating $\omega_{10}(t)$ that is generated in
the same way and with the same amplitude as in Figure \ref{fig:RTN}b.
We show the result in Fig.~\ref{fig:RTN_decay}.
By comparing Fig.~\ref{fig:RTN_decay} with
Fig.~2b, we note that adding level fluctuations reduces the Rabi amplitude
and renormalizes the Rabi frequency. The result in Fig.~\ref{fig:RTN_decay}
is closer to what is seen experimentally \cite{Martinis04}.

In a charge qubit, the qubit energy splitting is $\hbar I_o/2e$ when
biased at the degeneracy point \cite{QB1}. So for a charge qubit
critical current fluctuations are
related to fluctuations in the qubit energy splitting by
\begin{equation}
\frac{{\delta \omega _{10} }}{{\lb \omega _{10}\rb }} = \frac{{\delta I_o
}}{{\lb I_o \rb }}\mbox{~.}
\end{equation}
Comparing this to the analogous expression for a phase qubit
(Eq.~(\ref{eq:w10_I0})), it is obvious that
for the same amount of noise in the critical current
$\delta I_o/\lb I_o \rb$, the level fluctuations
$\delta\omega_{10}/\omega_{10}$ of a charge qubit seem to be less
than those of a phase qubit. This is because the phase qubit must
be biased with a current slightly less
than $\lb I_o\rb$, which magnifies the critical current noise typically by a
factor of 100. However, the area of the Josephson junctions in charge qubits
is much smaller than in phase qubits. As a result, $\lb I_o\rb$ is
smaller in the charge qubit. (In Ref.~\cite{QB1}, the charge qubit has $I_o
\approx 30 $ nA. In Ref.~\cite{Martinis04}, the phase qubit has $I_o \approx
11.6 $ $\mu$A.) Thus for a fixed value of $\delta I_o$, 
the charge qubit has a much larger relative fluctuation
$\delta I_o /\lb I_o \rb$. Therefore we believe that level fluctuations
remain a concern for charge qubits.

\begin{figure} 
\centerline{\psfig{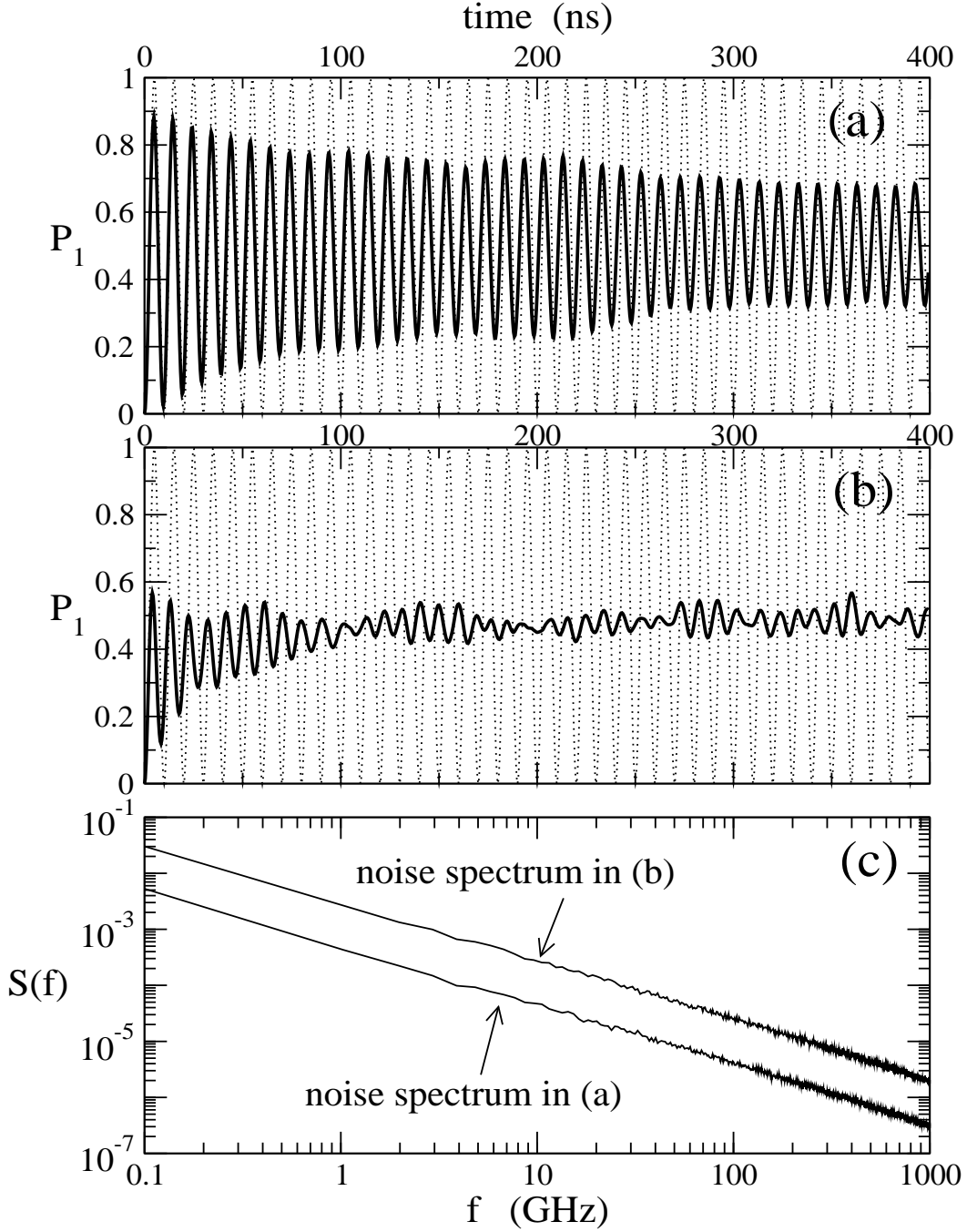}}
\caption{(a)-(b): Rabi oscillations in the presence of 1/f noise in the qubit
energy level splitting.
Dotted curves show the Rabi oscillations without the influence of noise. Panel
(c) shows the two noise power spectra $S(f) \equiv | \delta
\omega_{10}(f)/\langle \omega_{10} \rangle |^2$ of the fluctuations
in $\omega_{10}$ that were used to produce the solid curves
in panels (a) and (b). Rabi frequency $f_{R} = 0.1$ GHz.
 }\label{fig:1f} 
\end{figure}

1/f noise in the critical current of Josephson
junctions has been observed experimentally
\cite{vanHarlingen82,Foglietti86,Savo87,Wellstood88}.
The microscopic source of this noise is unknown, though
it has been suggested that the tunneling of atoms or ions is involved
\cite{Rogers85}.
When an ensemble of two level systems with a distribution
of relaxation times is considered, we can obtain the
resulting noise power spectrum by averaging the Lorentzian
power spectra of individual TLS over the distribution
of relaxation times \cite{Dutta}. For example we can replace $t_{TLS}$ in
the Lorentzian in
Eq. (\ref{eq:Lorentzian}) with the TLS relaxation
rate $\tau_{TLS}$ given by Eq. (\ref{eq:tau}), and average over
the TLS parameters $\Delta$ and $\Delta_{o}$ \cite{Yu04}.
The result is a 1/f noise spectrum in the critical current,
and hence in the qubit level fluctuations $\delta\omega_{10}$.
So we have considered the effect of 1/f noise in a Josephson junction
on qubit decoherence. We do not average over Lorentzians to obtain
our noise spectrum. Rather we numerically generate a time series
$\delta\omega_{10}$ with a 1/f noise power spectrum. We use this
$\delta\omega_{10}$ in Eq.~(\ref{eq:RTN}) and calculate the
Rabi oscillations with the same numerical approach as in Fig.~\ref{fig:RTN}.
The results are shown in
Fig.~\ref{fig:1f}. When the noise power is low, the Rabi amplitude is
reduced but it remains in phase with the unperturbed Rabi oscillations
(dotted line). However, as Fig.~\ref{fig:1f}b shows,
the coherent nature will eventually be destroyed as
the noise power
increases. The decaying Rabi oscillations clearly demonstrate that 1/f noise
is able to adversely affect the coherence of the qubit.

\section{Conclusion}

We have shown that the coupling between a two-level fluctuator and a
Josephson qubit has a large effect on the decay and decoherence
of the Rabi oscillations. We have studied the quantum
dynamics of a Josephson qubit subjected to microwave
driving by numerically integrating the time-dependent Schr\"odinger
equation. We focussed on two decoherence mechanisms of a Josephson
qubit caused by coupling to two level
systems. (A) In the resonant regime we considered a qubit in resonance with
a high-frequency TLS, i.e., a TLS with a large tunneling splitting.
Without any source of energy decay, the qubit occupation probability
$P_1$ of the excited state exhibits beating rather than being
sinusoidal as in the usual Rabi oscillations. Including the
energy decay of the excited TLS via phonon emission
results in the decoherence of the coupled
qubit-TLS system. This decoherence mechanism primarily causes a
characteristic relaxation time of the Rabi oscillations.
The Rabi dynamics at short times is not affected. Furthermore,
coupling between the TLS and
the microwave driving is found to further degrade qubit coherence. (B)
The other regime involved qubit level fluctuations.
Low-frequency TLS are treated as noise sources because they
randomly modulate the junction critical current $I_o$. Fluctuations in $I_o$
modulate the qubit energy splitting, thus randomizing the phase of the
qubit, leading to decoherence. Based on noise measurements in prior
experiments, our model calculations suggest that noise from a single TLS can
cause Rabi oscillations to decay within 100 ns. When the 
qubit is coupled to a single slow fluctuator, we have shown that the Rabi
decay time depends on the noise amplitude $\delta \omega_{10}$, the
characteristic fluctuation rate, and the Rabi frequency. When the qubit
level fluctuations have a $1/f$ noise spectrum, the Rabi oscillation
degradation increases with increasing noise power.

\section{acknowledgement}

We thank John Martinis, Robert McDermott and Ken Cooper for helpful
discussions. This work was supported in part by DOE grant
DE-FG02-04ER46107 and ONR grant N00014-04-1-0061.

\begin{thebibliography}{44}
\bibitem[{noi()}]{noise_T}
\bibinfo{note}{The thermally activated behavior $t_{TLF}^{-1} = (1/t_0) \exp
  (-E_a/k_B T)$ was experimentally verified in Ref.~\cite{Rogers84} for $T \geq
  15 K$. However since we are considering a much lower temperature, the above
  relation might not hold.}
\end{thebibliography}
\end{document}